# Interlayer coupling and gate-tunable excitons in transition metal dichalcogenide heterostructures

Shiyuan Gao [1], Li Yang [1, 2] and Catalin D. Spataru [3]*

1) Department of Physics, Washington University in St. Louis, St. Louis, Missouri 63136, USA
2) Institute of Materials Science and Engineering, Washington University in St. Louis, St. Louis, Missouri 63136, USA
3) Sandia National Laboratories, Livermore, California 94551, USA

*Corresponding author: cdspata@sandia.gov

ABSTRACT

Bilayer van der Waals (vdW) heterostructures such as $MoS_2/WS_2$ and $MoSe_2/WSe_2$ have attracted much attention recently, particularly because of their type II band alignments and the formation of interlayer exciton as the lowest-energy excitonic state. In this work, we calculate the electronic and optical properties of such heterostructures with the first-principles GW+Bethe-Salpeter Equation (BSE) method and reveal the important role of interlayer coupling in deciding the excited-state properties, including the band alignment and excitonic properties. Our calculation shows that due to the interlayer coupling, the low energy excitons can be widely tunable by a vertical gate field. In particular, the dipole oscillator strength and radiative lifetime of the lowest energy exciton in these bilayer heterostructures is varied by over an order of magnitude within a practical external gate field. We also build a simple model that captures the essential physics behind this tunability and allows the extension of the *ab initio* results to a large range of electric fields. Our work clarifies the physical picture of interlayer excitons in bilayer vdW heterostructures and predicts a wide range of gate-tunable excited-state properties of 2D optoelectronic devices.

**Keywords**: van der Walls heterostructure, interlayer exciton, transition metal dichalcogenides

Two-dimensional (2D) transition metal dichalcogenides (TMDC) and its heterostructures have attracted a lot of attentions recently as a promising candidate for photonics, optoelectronics, and valleytronics devices [1, 2]. With the type II band alignment, TMDC bilayer heterostructures possess ultrafast charge transfer and long-lived interlayer exciton as its lowest-energy optical excitation, which is desirable for light harvesting applications [3-5], as well as realizing high-temperature excitonic superfluidity [6]. In particular, because the interlayer charge transfer and exciton photoemission in these structures depends critically on the interlayer coupling [7-10] and the widely-used gate field can efficiently tune the band offset and interlayer interactions, the range of tunable optoelectronic properties of these heterostructures may be substantially enhanced than those of semiconductor quantum wells [11].

The first-principle density functional theory (DFT) + GW/ Bethe-Salpeter Equation (BSE) method has been very successful in studying and predicting excited-state properties of 2D structures [12-14]. However, previous calculations have only studied the band structure and excitonic properties of intrinsic TMDC heterostructures [15-17]. In these calculations, the role of interlayer coupling is not yet well addressed. Particularly, how the external gate field impacts the interlayer coupling and interlayer excitations is largely unknown. In this sense, it is essential to have a reliable study that can capture this interlayer coupling and excitons and how those electron-hole pairs and optical response are tuned by the external gate field.

In this work, we study gated $MoS_2$/$WS_2$ and $MoSe_2$/$WSe_2$ heterostructures with first-principles DFT + GW/ BSE calculations. We first study the band structure and interlayer coupling at the DFT level. Then we construct a simple model that accounts for the interlayer coupling and provides accurate results for the low-energy excitation spectrum both at the single-particle level and including electron-hole interactions. This allows us to predict the energy, dipole strength, and radiative lifetime of the excitons under arbitrary external field, with input parameters obtained from the first-principle results. As we will show, apart from the obvious linear response of the band alignment to the external field, the interlayer coupling between the valance bands of the bilayer leads to anti-crossing behavior of the lowest energy excitons, changing its nature continuously from interlayer to intralayer. Our result reveals the nature of the excitons in TMDC heterostructures and explains their gate tunability, which will help to interpret and predict experimental optical measurement.

In the following, we use $MoS_2$/$WS_2$ as the primary example. Figure 1 shows a simple schematic picture of the band alignment $MoS_2$/$WS_2$ bilayer heterostructure based on our calculation. The conduction band minimum (CBM) and valence band maximum (VBM) belongs to $MoS_2$ and $WS_2$ respectively, forming a type II band alignment. Spin-orbit coupling (SOC) further splits the spin-up and spin-down bands. Finally, due to the band alignment, the lowest optical transition (exciton) is interlayer, while the intralayer transitions lays higher up in energy. To further quantify this picture, we turn to the first-principles calculation, beginning with the DFT picture.

The DFT calculation is done with plane-wave pseudopotential method implemented in Quantum Espresso [18], using the PBE exchange-correlation functional [19] and including the semi-core states of Mo and W. The plane-wave cutoff is chosen to be 75 Ry to ensure the converged results. The structure is relaxed with the van der Waals (vdW) DFT-D2 functional [20, 21]. A vacuum of at least 15 Å is added in the vertical direction to avoid spurious interactions between adjacent slabs. We adopt the stable AB-stacking configuration between the two layers, which can be achieved experimentally by epitaxial growth [22]. The interlayer coupling strength may change with a different stacking configuration or twisting angle [23], but the physical picture and the model we are about to present remains the same. The relaxed lattice constant and interlayer distance (vertical distance from Mo to W) is 3.18 Å and 6.21 Å for $MoS_2/WS_2$, and 3.32 Å and 6.54 Å for $MoSe_2/WSe_2$, which agrees with previous calculations [24, 25]. (The lattice mismatch between the two layers is less than 0.5 %, which will not significantly affect the calculated results.) A saw-tooth like potential in the vertical direction is used to simulate the external gating field along with the dipole correction. The positive direction of the external field is defined as pointing from the $MoS_2$ layer towards the $WS_2$ layer (see Figure 2(b)). While a gate field usually leads to charging of the material, electrostatic doping effects [26-29] can be avoided in experimental set-ups where the sample is not directly contacted by metal electrodes.

Figure 2(a) shows the DFT band structure of $MoS_2/WS_2$ without external field, with the color indicating the projection of the wavefunction onto either layer. To simplify the problem, SOC is not included here for the sake of argument, but will be included later in the final result. As expected, the DFT result confirms the Type II band alignment, where the VBM at the K point is in the $WS_2$ layer and the CBM is at the Γ point in the $MoS_2$ layer. However, the projection of the electronic wavefunction shows that the VBM wavefunction is not 100% $WS_2$, but rather have a layer projection of 90% $WS_2$ and 10% $MoS_2$, indicating the presence of interlayer coupling. This is more evident by looking at how the VBM and CBM energy changes in response to a vertical electric field, as indicated by dots in Figure 2(c). Because of the interlayer coupling, the two bands at VBM shows anti-crossing behavior as the electric field reverses their order. In contrast to this, due to the lack of interlayer interaction, the two bands at CBM pass through each other. This coupling behavior can be described by a simple model of a 2×2 matrix $\begin{pmatrix} \varepsilon_{Mo} & t \\ t & \varepsilon_W \end{pmatrix}$, where $\varepsilon_{Mo}$ and $\varepsilon_W$ are the energy of the VBM of each layer alone. Under external electric field $E$, the relative band alignment of the two layers responds as $\varepsilon_{Mo}(E) - \varepsilon_W(E) = \varepsilon_{Mo}(0) - \varepsilon_W(0) - eEd/\epsilon$, where d is the interlayer separation and the parameter $\epsilon$ reflects the material's screening response to the vertical electric field. For the MoS$_2$/WS$_2$ bilayer we find $\epsilon = 6.5$, in agreement with the fully *ab initio* calculated dielectric constant of bulk MoS$_2$ and WS$_2$ (~7) [30]. The same value for $\epsilon$ also describes the response of CBM energy to the electric field, where the coupling is negligible. With a value of $t = 45\ meV$ for the parameter describing the interlayer coupling strength at the VBM, this simple model perfectly describes the first-principles result as demonstrated in Fig. 2(c).

In addition to the single-particle level, the interlayer coupling also plays a critical role in determining the properties of the heterostructure at the excitonic level. In order to show that, a many-body GW+BSE calculation is performed, which is proven to be reliable in calculating excited-state properties of monolayer TMDCs [14, 31]. The dielectric function is evaluated with an 18×18×1 k-point grid, 10 Ry energy cutoff and 400 conduction bands in the summation. The quasiparticle band gap is then determined from a single-shot G$_0$W$_0$ calculation. The excitonic effects are included by solving the Bethe-Salpeter Equation (BSE) on a finer k-point grid of 90×90×1. A slab Coulomb truncation is implemented to avoid interactions between periodic images [32]. These calculations are done with the BerkeleyGW package [33].

For MoS$_2$/WS$_2$, the GW correction increases the direct band gap at K from 1.62 eV to 2.42 eV. A scissor operator is used to fit the GW correction to the first two valence and conduction bands near the K point, which introduces no more than 0.01 eV error compared with the full GW result. Our further calculations show that the GW correction is insensitive to the external electric field.

At the BSE level, we determine the energy and dipole oscillator strength of the interlayer exciton, as shown by the dots in Figure 3. In the following, we reveal the key role played by the coupling between the valence bands in explaining these results. Focusing on the lowest-energy interlayer and intralayer excitons, i.e. the so-called "1s" exciton, the BSE can be written in the basis of individual (uncoupled) layers as

$$(\varepsilon_c - \varepsilon_v)A_{vc}^S + \sum_{v'c'}\langle vc|K|v'c'\rangle A_{v'c'}^S = \Omega^S A_{vc}^S, \tag{1}$$

where $|S\rangle = \sum_{vc} A_{vc}^S |vc\rangle$ is the exciton eigenstate, $v$ and $c$ are the layer index of the valence and conduction band, $\varepsilon_v$ and $\varepsilon_c$ are the bare band edge energies of each layer (without including the interlayer coupling), $\Omega^S$ is the exciton energy, and $K$ is the electron-hole interaction kernel. We have absorbed the k-point indices and focus on the effect of interlayer coupling. The non-zero matrix elements of the kernel $K$ are the binding energies of the exciton between two individual layers (MoS$_2$ intralayer, WS$_2$ intralayer, and interlayer exciton, respectively):

$$\langle v_{Mo} c_{Mo} | K | v_{Mo} c_{Mo} \rangle = -\varepsilon_{B,Mo},$$

$$\langle v_W c_W | K | v_W c_W \rangle = -\varepsilon_{B,W},$$

$$\langle v_W c_{Mo} | K | v_W c_{Mo} \rangle = \langle v_{Mo} c_W | K | v_{Mo} c_W \rangle = -\varepsilon_{B,inter},$$

which are treated as parameters in the model and are extracted from the *ab initio* calculation. It's worth noting that the intralayer exciton binding energy here should be differentiated from that of an isolated monolayer, because although it assumes no interlayer hopping, it does reflect the impact of dielectric screening from the other layer.

Combining the above assumptions, the BSE Hamiltonian can be written in the bare electron-hole basis as a 4-by-4 matrix:

$$\begin{pmatrix} \varepsilon_{c_{Mo}} - \varepsilon_{v_W} - \varepsilon_{B,inter} & t & & \\ t & \varepsilon_{c_{Mo}} - \varepsilon_{v_{Mo}} - \varepsilon_{B,Mo} & & \\ & & \varepsilon_{c_W} - \varepsilon_{v_W} - \varepsilon_{B,W} & t \\ & & t & \varepsilon_{c_W} - \varepsilon_{v_{Mo}} - \varepsilon_{B,inter} \end{pmatrix} \begin{matrix} (\leftrightarrow v_W c_{Mo}) \\ (\leftrightarrow v_{Mo} c_{Mo}) \\ (\leftrightarrow v_W c_W) \\ (\leftrightarrow v_{Mo} c_W) \end{matrix}$$

The eigenvalues and eigenvectors of this matrix describe the energy and layer composition of the exciton eigenstates. The off-diagonal interlayer coupling $t$ is responsible for mixing the intralayer and interlayer exciton. This simple model can reproduce the exciton energy and dipole strength from the *ab initio* calculation very well, as shown in Figure 3, thus validating the assumptions made above. The parameters used in this model for MoS2/WS2 and MoSe2/WSe2 heterostructures are summarized in Table I.

Without interlayer coupling, the interlayer exciton energy would change linearly with the electric field E via the (quantum-confined) Stark effect [11, 34], while its oscillator strength would be independent of E. However, as shown in Figure 3(a), due to the interlayer coupling, the interlayer and MoS$_2$ intralayer exciton states mix leading to an anti-crossing behavior. The electric field can gradually tune the nature of the lowest-energy exciton of the heterostructure from primarily interlayer to primarily intralayer. During this process, the overlap of the electron and hole wavefunction gradually increase, resulting in a gradually enhanced exciton oscillator strength, as shown in Figure 3(b). This model demonstrates how the interlayer coupling enables the tuning of excitonic properties with external electric field.

Having benchmarked our model against first principles calculations in the absence of SOC, we now proceed to add this additional effect which splits the spin-up and spin-down bands and changes the exciton energy, as shown in the schematic plot of Fig. 1. Because spin along the vertical direction is a good quantum number near the K point [35], we can include SOC as a perturbation. We take the spin-orbit correction to be the band energy difference between a noncollinear spin-orbit DFT calculation and a spin-unpolarized DFT calculation without interlayer coupling: $\Delta\varepsilon_{nk\sigma}^{SO} = \varepsilon_{nk\sigma}^{SO} - \varepsilon_{nk}$. The spin-orbit correction to the GW quasiparticle energies is assumed to be the same as the spin-orbit correction to the DFT energies [36]. The spin-orbit splitting at the VBM is around 160 meV for MoS$_2$ and 440 meV for WS$_2$, while at the CBM it is only 3 meV for MoS$_2$ and 40 meV for WS$_2$. The interlayer interaction couples the like-spin bands the same way as discussed before. Therefore, the band anti-crossing like the one in Fig. 2(c) is still present, but only with valence bands of like-spin (see SI).

Following Ref. [36], the spin-orbit energy correction for an exciton eigenstate between two individual layers, as defined in Eq. (2), is $\Delta\Omega_{vc\sigma}^{S} = \langle vc|H_{\sigma}^{SO}|vc\rangle = \sum_{\boldsymbol{k}}|\tilde{A}_{vc\boldsymbol{k}}^{S}|^{2}(\Delta\varepsilon_{c\boldsymbol{k}\sigma}^{SO} - \Delta\varepsilon_{v\boldsymbol{k}\sigma}^{SO})$. We've neglected the exchange part of the electron-hole interaction, which is below 20 meV, much smaller than the SOC splitting. Then the full exciton eigenstate is solved with the interlayer coupling following the same procedure. Finally, the imaginary part of the dielectric function is calculated using formula $\epsilon_2(\omega) = \frac{16\pi^2 e^2}{\omega^2}\sum_{S\sigma}|\boldsymbol{e}\cdot\langle 0|\boldsymbol{v}|S\sigma\rangle|^2\delta(\omega - \Omega_\sigma^S)$, where $\boldsymbol{e}$ is the polarization of the incident light, $\boldsymbol{v}$ is the velocity operator and $|S\sigma\rangle$ is the exciton eigenstate with spin $\sigma$. Higher excitonic states in the series such as the 2s state are not included.

The calculated exciton energies together with the simulated absorption spectrum $\epsilon_2(\omega)$ at different electric fields are shown in Figure 4(a) and 4(b) for MoS$_2$/WS$_2$ and MoSe$_2$/WSe$_2$, respectively. After including SOC, the spin-up bands (associated exciton states indicated by blue dashed lines) of WX$_2$ and the spin-down bands (associated exciton states indicated by red dashed lines) of MoX$_2$ (X=S, Se) have the higher energy at the K point. The top valence band, mainly from WX$_2$ and responsible for the lowest energy interlayer exciton, moves further apart from the like-spin band from MoX$_2$. Therefore, for the lowest energy exciton, the impact of the interlayer coupling is weaker at zero field with the inclusion of SOC and the anti-crossing behavior is apparent only for |E| > 5V/nm for MoS$_2$/WS$_2$, resulting in a more linear Stark shift and a smaller oscillator strength of lowest exciton at low field. On the other hand, the anti-crossing behavior is seen in the higher-lying exciton states at relatively low electric field values as shown in Fig. 4(a), which provides an easy way to experimentally determine the strength of the interlayer coupling by optical absorption measurements.

The MoSe$_2$/WSe$_2$ bilayer heterostructure shares a very similar band alignment (see SI) and interlayer coupling with MoS$_2$/WS$_2$ (see Table I). Therefore, as shown in Fig. 4(b), the dependence of exciton energy and optical absorption on the electric field in MoSe$_2$/WSe$_2$, including the anti-crossing behavior, is very similar to that of MoS$_2$/WS$_2$, with only quantitative difference. Specifically, the exciton energies are about 0.2 eV lower in MoSe$_2$/WSe$_2$. Moreover, due to slightly different band energy and SOC, the anti-crossing point for the lowest exciton is moved to an even higher electric field around 6V/nm.

Although the anti-crossing behavior in energy is inaccessible at low electric field for the lowest exciton, the impact of the interlayer coupling can still be seen from the tunability of the radiative lifetime. The radiative lifetime of the lowest exciton is an important parameter for photoluminescence and electron-hole separation process. For 2D materials, the intrinsic radiative lifetime of exciton at zero temperature is directly related to the dipole strength by $\tau_S^{-1} = \frac{8\pi e^2 \Omega_S \mu_S^2}{\hbar^2 c A_{uc}}$, where $\Omega_S$ is the energy of the exciton state S, $\mu_S^2$ is the modulus square dipole strength of the exciton divided by the number of k-points, and $A_{uc}$ is the area of the unit cell. Then the exciton lifetime at finite temperature can be obtained by thermally averaging the exciton lifetime assuming a parabolic dispersion [15, 37]. Figure 4(c) and 4(d) shows the 0K and 300K radiative lifetime of the two lowest exciton branches. The lifetime value at zero electric field is in good agreement with a previous calculation [15]. Because of the electric field tuning of the exciton oscillator strength, the lifetime of the lowest exciton increases by two orders of magnitude as the electric field increase from -6V/nm to 4V/nm. This shows that the lifetime of lowest (interlayer) exciton in these heterostructures can be widely tuned by the external gate field.

Finally, it is worth noting that the same physical picture can also lead to gate-tunable excitons in bilayer homojunctions. We notice a recent experimental work has reported the observation of gate-tunable exciton energy and lifetime in bilayer WSe$_2$ [38].

In conclusion, we have studied the band alignment and excitonic properties of MoS$_2$/WS$_2$ and MoSe$_2$/WSe$_2$ bilayer heterostructures from first principles DFT and GW+BSE calculation. We have shown that interlayer coupling is the key to understanding their properties, allowing the nature of the lowest-energy exciton to be tuned gradually from interlayer to intralayer by an external gate field. This is accurately captured by our simple model which accounts for the interlayer coupling in the presence of electron-hole interactions, which predicts an anti-crossing behavior of the exciton energy, as well as widely tunable dipole oscillator strength and radiative lifetime of the lowest-energy excitons by an order of magnitude with an external gate field of a few V/nm. Our result provides a quantitative physical picture of excitons in bilayer vdW heterostructures which would benefit future investigations of the gate-tunable excited-state properties in 2D heterostructures.

## AUTHOR INFORMATION

The authors declare no competing financial interest.

## ACKNOWLEDGMENT

This work is supported by the Laboratory Directed Research and Development program at Sandia National Laboratories. Sandia National Laboratories is a multimission laboratory managed and operated by National Technology and Engineering Solutions of Sandia, LLC., a wholly owned subsidiary of Honeywell International, Inc., for the U.S. Department of Energy's National Nuclear Security Administration under contract DE-NA-0003525. S.G. and L.Y. are supported by the National Science Foundation (NSF) CAREER Grant DMR-1455346. S.G. also thanks James Bartz for the internship opportunity at SNL. The computational resources have been provided by the Stampede of Teragrid at the Texas Advanced Computing Center (TACC) through XSEDE.

**Figures:**

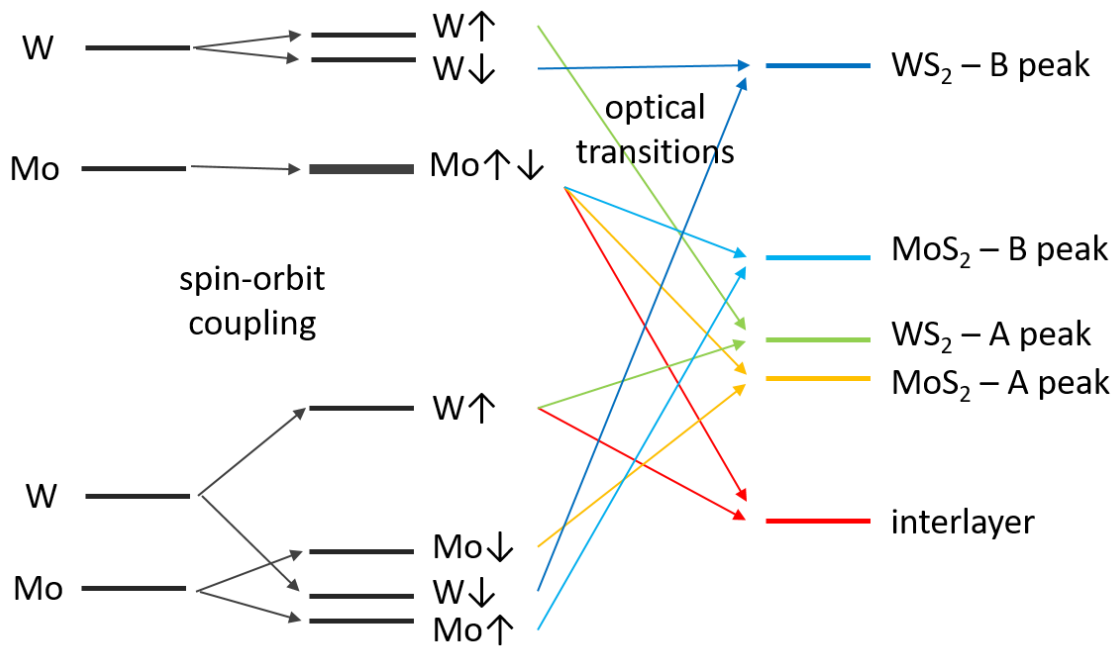

Figure 1 Schematic energy level diagram of $MoS_2/WS_2$ heterostructure showing its relative band alignment and related optical transitions without the interlayer coupling.

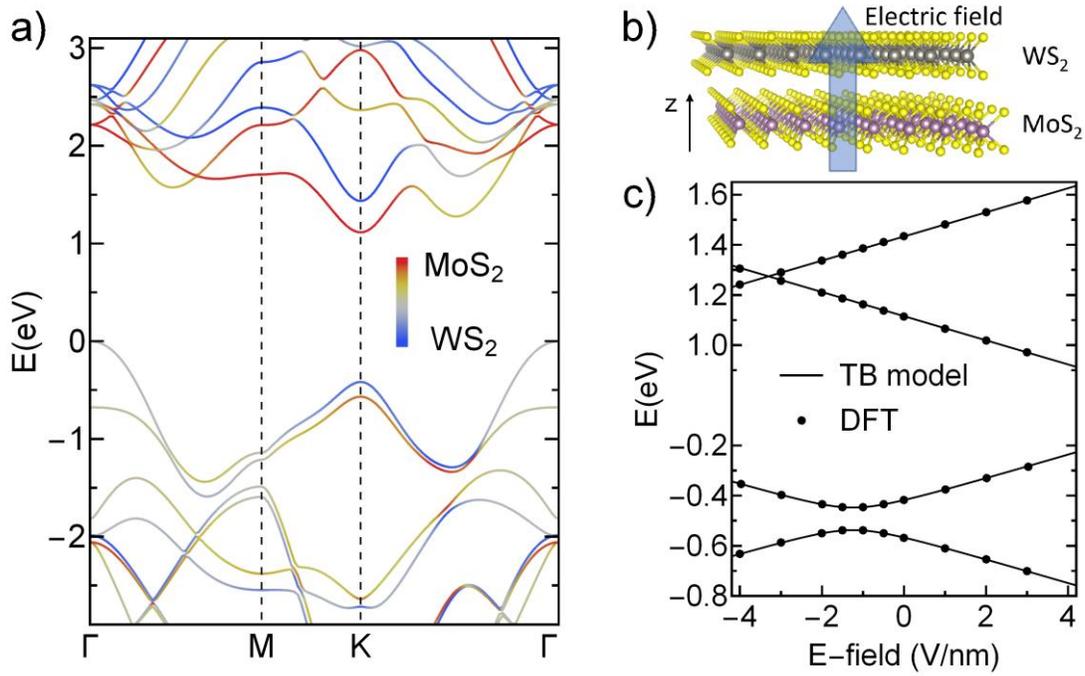

Figure 2 (a) DFT band structure of $MoS_2/WS_2$ heterostructure with the color indicating the projection of the wavefunction onto each layer. (b) Schematic plot showing an external electric field applied to the heterostructure. (c) The energy of the top two valence bands and bottom two conduction bands at K point as a function of the external electric field. All the results are obtained in the absence of spin-orbit coupling.

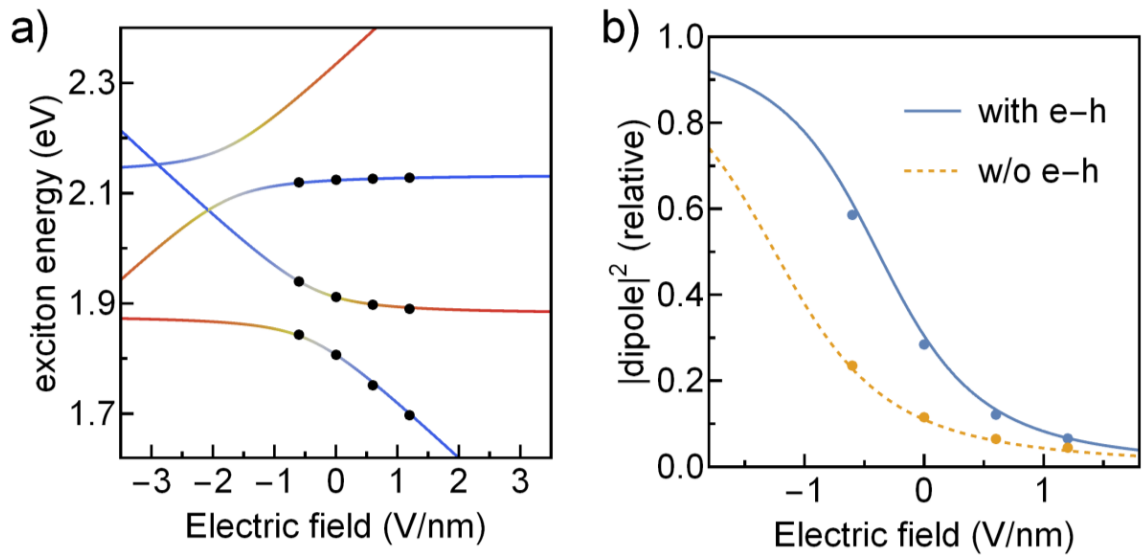

Figure 3 (a) Energy of the exciton eigenstates in $MoS_2/WS_2$ heterostructure as a function of the external electric field. Dots indicate the first-principles result and the line comes from the model. Color of the line indicate the character of the hole in the exciton. Red means the hole is in $MoS_2$ and blue is in $WS_2$.

(b) Squared transition dipole of the lowest energy exciton as a function of the external electric field, relative to the squared transition dipole of the lowest energy exciton in single layer MoS2 (for which we find a value –in atomic units- of $0.02$ per unit area). Dash line shows similar result without considering electron-hole interaction. The results are obtained in the absence of spin-orbit coupling.

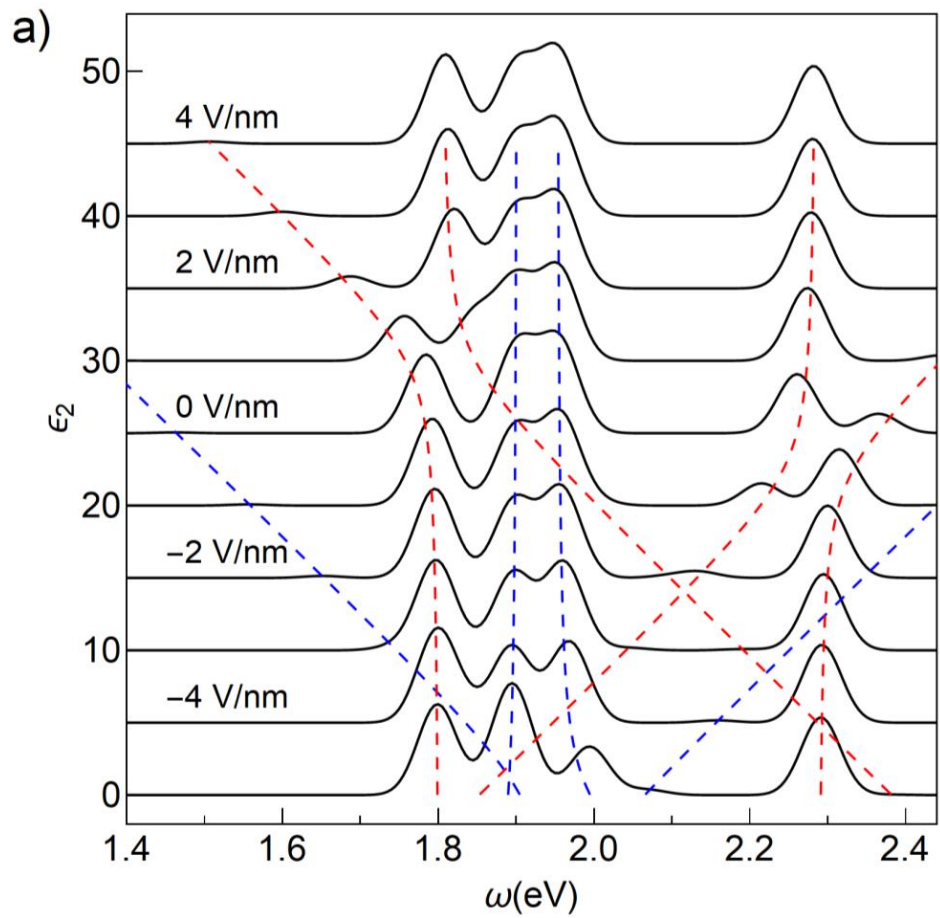

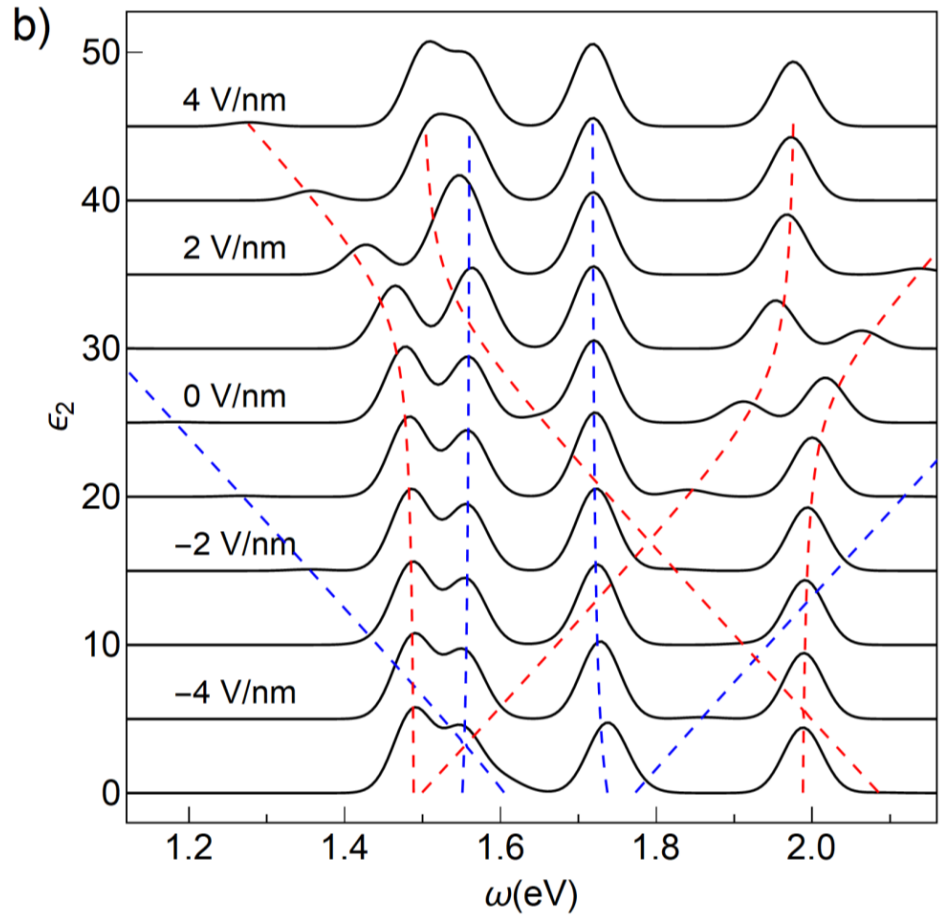

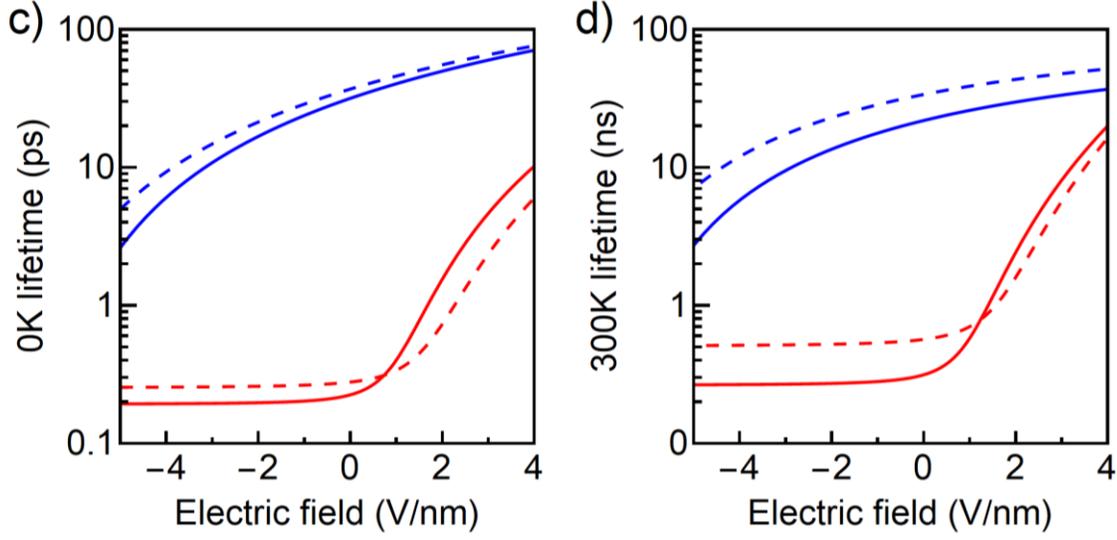

Figure 4 (a, b) Imaginary part of dielectric function $\epsilon_2$ for MoS$_2$/WS$_2$ (a) and MoSe$_2$/WSe$_2$ (b) heterostructure calculated from the model including spin-orbit coupling for different electric field values. Blue and red dashed line indicate the energy of the exciton with electron spin up and down, respectively. (c, d) Radiative lifetime of the lowest two exciton branches, with blue and red corresponding to their electron spin, for MoS$_2$/WS$_2$ (solid line) and MoSe$_2$/WSe$_2$ (dash line) heterostructure at 0K (c) and 300K (d) as a function of the external electric field.

| Heterostructure | $\epsilon$ | $t$ | $\varepsilon_{B,Mo}$ | $\varepsilon_{B,W}$ | $\varepsilon_{B,inter}$ |
|---|---|---|---|---|---|
| MoS$_2$/WS$_2$ | 6.5 | 45 meV | 0.6 eV | 0.57 eV | 0.51 eV |
| MoSe$_2$/WSe$_2$ | 7.4 | 49 meV | 0.56 eV | 0.53 eV | 0.49 eV |

Table I. Key parameters in the model for MoS$_2$/WS$_2$ and MoSe$_2$/WSe$_2$ heterostructure.

**TOC Figure**

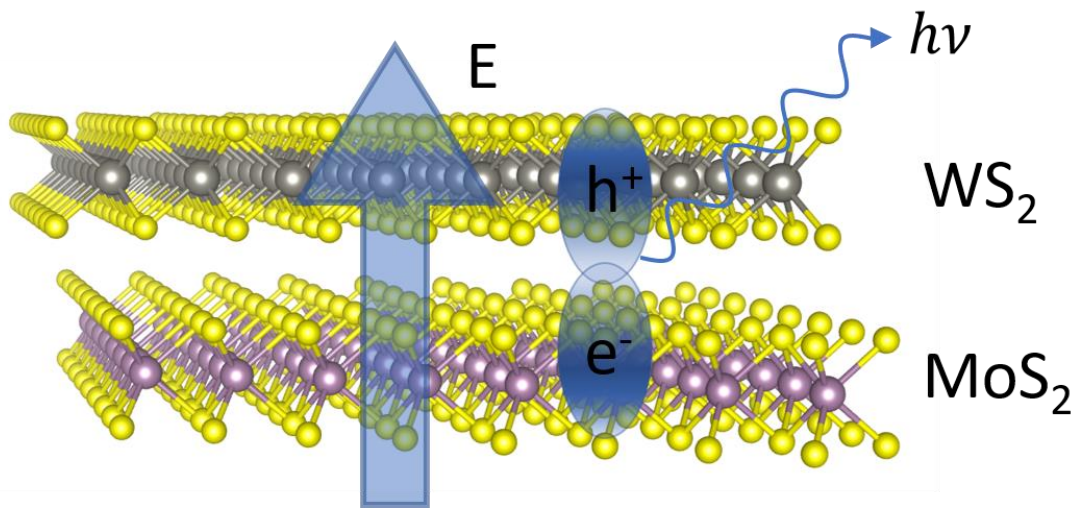

**Supplementary Information**

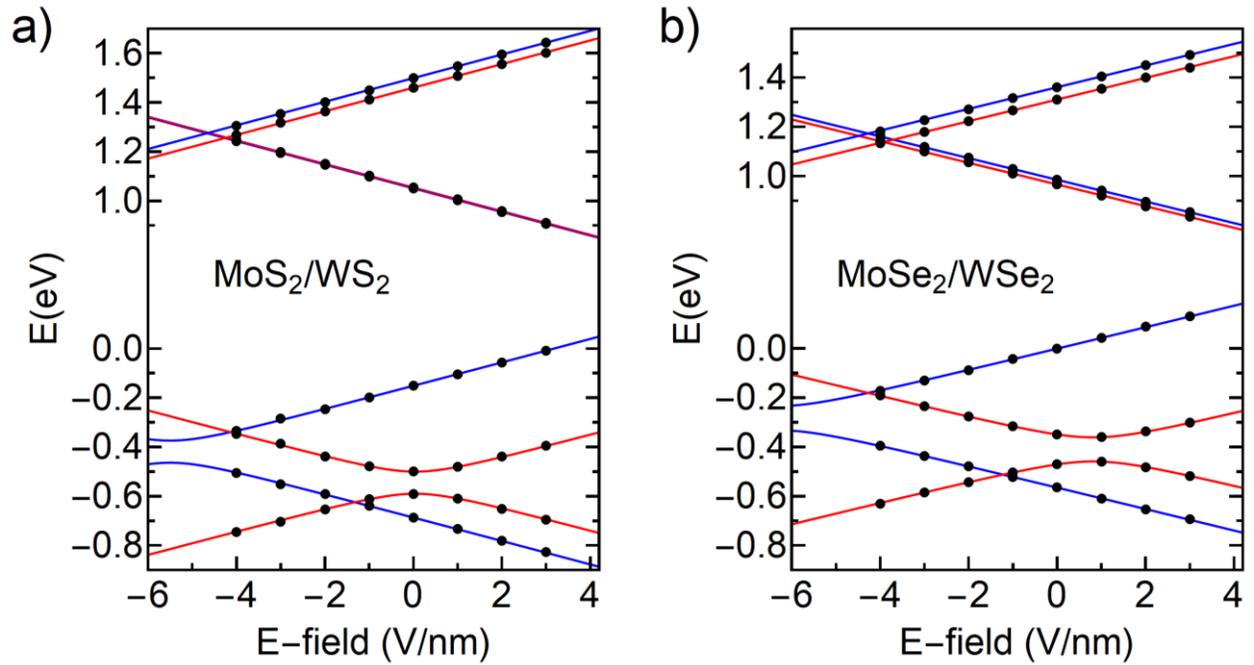

Figure S1 The energy of the top valence bands and bottom conduction bands at K point versus the external electric field for $MoS_2/WS_2$ (left) and $MoSe_2/WSe_2$ (right) heterostructure including the spin-orbit coupling. The dotted are the DFT-calculated results and the lines are from our interlayer coupling model. The blue and red lines represent spin-up and spin-down bands, respectively.